\documentclass[prl,aps,twocolumn,amsmath,amssymb,floatfix]{revtex4}

\usepackage{graphicx}
\usepackage{dcolumn}
\usepackage{bm}

\begin{document}

\preprint{APS/123-QED}

\title{Excitation spectrum of two correlated electrons in a lateral quantum dot \\
with negligible Zeeman splitting}

\author{C. Ellenberger,$^1$ T. Ihn,$^1$ C. Yannouleas,$^2$ U. Landman,$^2$ 
K. Ensslin,$^1$ D. Driscoll,$^3$ and A.C. Gossard$^3$}
\affiliation{
$^1$Solid State Physics, ETH Zurich, 8093 Zurich, Switzerland\\
$^2$School of Physics, Georgia Instititute of Technology, Atlanta, Georgia 30332-0430\\
$^3$Materials Department, University of California, Santa Barbara, CA 93106
}

\date{\today}

\begin{abstract}
The excitation spectrum of a two-electron quantum dot is investigated by 
tunneling spectroscopy in conjuction with theoretical calculations. The dot made from a material with 
negligible Zeeman splitting has a moderate spatial anisotropy leading to a 
splitting of the two lowest triplet states at zero magnetic field. In addition 
to the well-known triplet excitation at zero magnetic field, two additional 
excited states are found at finite magnetic field. The lower one is identified 
as the second excited singlet state on the basis of an avoided crossing with the
first excited singlet state at finite fields. The measured spectra are in 
remarkable agreement with exact diagonalization calculations. The results prove 
the significance of electron correlations and suggest the formation of a state 
with Wigner-molecular properties at low magnetic fields.
\end{abstract}

\pacs{Valid PACS appear here}
\maketitle


Two-electron (2$e$) systems (e.g., the He atom \cite{Tanner00} and the H$_2$ molecule 
\cite{ref2}) have played an important role in the development of the theory of 
many-body effects in quantum mechanics and the understanding of the
chemical bond. Recently, two-dimensional (2D) man-made 
quantum dots (QDs), fabricated in semiconductor materials have attracted
attention as a laboratory for investigations
of few-body systems with highly controlled parameters 
\cite{Meurer92,Ashoori93,Schmidt95,Tarucha96,Kouwenhoven97,Ciorga00,%
Kyriakidis02,Zumbuhl04} such as the electron number, confinement strength, 
interelectron repulsion, and the influence of an applied magnetic field $(B)$.
Furthermore, 2$e$ QDs may be used for implementing
logic gates in quantum computing \cite{Loss98}.
 
Earlier experiments on vertical QDs led to the observation of a singlet--triplet (ST) 
transition in the 2$e$ ground state as a function of the 
magnetic field \cite{Ashoori93,Schmidt95,Kouwenhoven97} and for $N \geq 2$ to an
energy level shell structure reminiscent of the periodic table of natural atoms 
\cite{Tarucha96}. The spin-triplet excited state of the 2$e$ system was 
investigated in Ref.~\onlinecite{Kouwenhoven97} in finite bias tunneling experiments.
Laterally defined few-electron dots \cite{Ciorga00} added 
investigations of the magnetic-field dependent ST gap for 
two electrons \cite{Kyriakidis02,Zumbuhl04}. 

At low magnetic fields, the importance of electron correlations in 2D QDs (under
appropriate conditions, see below), with different behavior compared to
natural atoms, has been anticipated theoretically 
\cite{Wagner92,pfann,yann99,Egger99,Yannouleas00,mikh,yann7}.
Clear experimental evidence for such strong correlation effects
is still lacking. Insights can be gained through investigations of the excitation spectrum of 
$2e$ QDs in a perpendicular magnetic field. Here
we report on joined experimental and theoretical endeavors that establish the
correlated nature of the two electrons in our lateral QD, purposely
fabricated to exhibit a negligible Zeeman splitting. We find that the
explanation of the measured spectra requires a high level treatment of
electron correlations in the dot, and cannot be accounted for by perturbative
schemes starting from the independent-particle picture. The exact-diagonalization
(EXD) calculations are in remarkable agreement with intricate measured features 
of the spectra, indicating the formation of an H$_2$-type Wigner molecule (WM) 
at low magnetic field, lifting of degeneracies, and the appearance of avoided 
crossings between excited singlet states associated with a shape anisotropy of 
our dot -- as well as excitations resulting from relative and center-of-mass 
(CM) motion.

\begin{figure}[t]
\centering
\includegraphics[width=6.5cm]{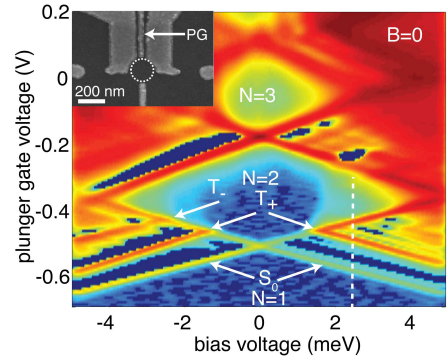}
\caption{(Color online) Differential conductance $dI/dV_\mathrm{bias}$ at zero magnetic field. Electron numbers $N$ are indicated in the diamonds. The arrows mark the transitions between the one-electron ground state and the 2$e$ spin-singlet ($S_0$) and spin-triplet state ($T_1$). Inset: scanning electron micrograph of the sample. The quantum dot is in marked with a dotted circle and the gate PG is the plunger gate. The vertical dashed line marks the bias voltage for which the data in Fig.~\ref{figure2} is taken.\label{figure1}}
\end{figure}

The sample is based on a 55~nm wide Ga[Al]As parabolic quantum well \cite{Sundaram88} with the center located 75~nm below the surface. The parabola ranges from 6\% Al-concentration in the center to 23\% at the edges. It was designed for having an average $g$-factor for electrons $|g|$$\approx$0 \cite{Salis01}. The effective mass of the electrons is higher than that of pure GaAs due to the larger band gap. We use the estimated value $m^\star=0.07m_0$, leading to an effective Rydberg energy $R_\mathrm{Ry}^\star=6$~meV. The two-dimensional electron gas (2DEG) has a density $n_\mathrm{s}=4.5\times 10^{15}$~m$^{-2}$ and a mobility $\mu=2.5$~m$^2$/Vs at a temperature of 4.2~K. A highly Si-doped layer 1.76~$\mu$m below the surface acting as a backgate was not used during the experiments described below.
The quantum dot (geometric diameter 220~nm) with integrated charge readout depicted in Fig.~\ref{figure1} (inset) was defined using electron beam lithography patterned split-gates. All measurements were performed at an electronic temperature below 300~mK as estimated from the width of conductance resonances in the Coulomb-blockade regime.

Figure~\ref{figure1} shows the differential conductance of the QD in the plane of bias and plunger-gate voltages. It was measured by applying the DC voltage $V_\mathrm{bias}/2$ to the source and $-V_\mathrm{bias}/2$ to the drain contact. The recorded current $I$ was numerically differentiated in order to obtain the differential conductance $dI/dV_\mathrm{bias}$. The resulting diamond pattern of suppressed conductance corresponds to fixed electron numbers $N$ in the dot as indicated in the figure.

The values for the electron numbers were found by depleting the dot down to zero electrons with more negative gate voltages (not shown) and by using the quantum point contact near the quantum dot as a charge detector.
For these modified gate voltage settings we observed the first single-particle excited state of the $N=1$ quantum dot at an energy $\Delta_{1}= 5$~meV above the ground state and found an addition energy $\Delta\mu_2=6.9$~meV (both $\Delta_1$
and $\Delta \mu_2$ are outside the range shown in Fig.\ 1).  These numbers show 
that in this dot, confinement effects ($\Delta_1$) and interaction effects 
($\Delta \mu_2$) are comparable in magnitude.

Returning to Fig.~\ref{figure1}, we determine from the extent of the $N=2$ 
diamond in bias direction the addition energy $\Delta\mu_3=5.1$~meV for adding the third electron to the dot.
The strong lines labeled $S_0$ in the figure correspond to tunneling transitions between the one-electron ground state and the 2$e$ spin-singlet ground state.

Strong lines parallel to the diamond boundaries outside the diamonds correspond to excited states of the level spectrum. In Fig.~\ref{figure1}, we see the transition $T_+$ between the one-electron ground state and the first excited 2$e$ triplet state. The energy separation between the singlet and this triplet state is $J=1.7$~meV. For negative bias voltage we see an additional excited state $T_-$ which we attribute to the second triplet state split from $T_+$ by 1.5~meV due to 
the deviation of the dot potential from 
circular shape (see also Fig.~\ref{figure4} below).
Negative differential conductance can arise due to asymmetric coupling of states to source and drain. Inside the diamonds,
higher order (cotunneling) processes can be seen for $N\geq 2$ \cite{deFranceschi01}. 

\begin{figure}[tbhp]
\centering
\includegraphics[width=6.5cm]{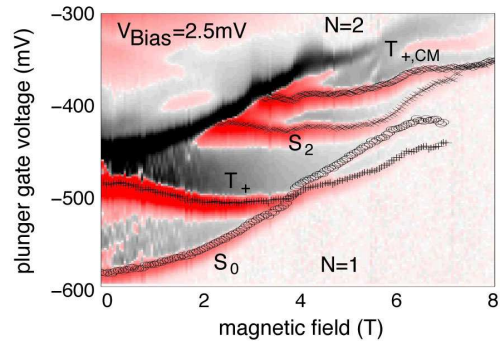}
\caption{(Color online) Differentiated current $dI/dV_\mathrm{pg}$ at $V_\mathrm{bias}=2.5$~mV. Red regions correspond to negative $dI/dV_\mathrm{pg}$, black to positive values. Electron numbers $N$ are indicated. Transitions between the one-electron ground state and the 2$e$ spin-singlet ground state ($S_0$), spin-triplet excited state ($T_+$) and spin-singlet excited state ($S_2$) are labeled.
\label{figure2}
}
\end{figure}

In order to investigate the evolution of the excited state spectrum with 
magnetic field, we have measured the current at fixed $V_\mathrm{bias}=2.5$~mV. In Fig.~\ref{figure2} we plot the derivative $dI/dV_\mathrm{pg}$. Resonances in $dI/dV_\mathrm{pg}$ correspond to resonances in the differential conductance $dI/dV_\mathrm{bias}$.
The resonances indicated by symbols in the plot are labeled consistently with 
Fig.~\ref{figure1}. It can be seen that the ST separation 
$J(B)=E_{T_+}-E_{S_0}$ between the $S_0$ and the $T_+$ states decreases with magnetic field and becomes zero at $B_\mathrm{ST}\approx 4$~T. Beyond this field, 
the energy of the $S_0$ state continues to increase and exhibits an avoided crossing with another excited state $S_2$ which appears above 2.7~T. The avoided crossing suggests that $S_2$ is an excited spin-singlet state and that the deviation of the dot potential from axial symmetry causes level repulsion. 
A third excited state ($T_\mathrm{+,CM}$) appears at even higher plunger gate voltage.
This state is shown below [see Fig.\ 4] to be a triplet excited state 
associated with the CM motion of the $2e$ molecule. 

A qualitative indication of the importance of correlation effects
can be obtained via the Wigner-parameter 
$R_\mathrm{W} = (e^2/\kappa l_0)/\hbar\omega_0$,
i.e., the ratio of the Coulomb interaction for two electrons at distance 
$l_0=\sqrt{\hbar/m^\star\omega_0}$ and the single-particle level spacing in a 
circular harmonic confinement, $\hbar\omega_0$ \cite{yann99}. Correlation 
effects are of increasing importance when the interaction energy becomes 
dominant over the single-particle spacing, i.e., when $R_\mathrm{W}>1$. Using a 
conservative estimate for the confinement energy of 5~meV (as quoted above for 
the dot with zero or one electrons), we find 
$R_\mathrm{W}\approx 1.55$ indicating that correlations may indeed have a
certain significance.

Quantitative understanding of the measured excitation spectrum and of the 
underlying strong correlations in the $2e$ QD were obtained through exact
diagonalization of the Hamiltonian for two 2D interacting electrons
\begin{equation}
{\cal H} = H({\bf r}_1)+H({\bf r}_2)+\gamma e^2/(\kappa r_{12}),
\label{ham}
\end{equation}
where the last term is the Coulomb repulsion, $\kappa$ (12.5 for GaAs) is the
dielectric constant, and
$r_{12} = |{\bf r}_1 - {\bf r}_2|$. 
The prefactor $\gamma$ accounts for the reduction of the Coulomb strength
due to the finite thickness of the electron layer in the $z$-direction
and for any additional screening effects due to the gate electrons.  
$H({\bf r})$ is the single-particle Hamiltonian, which 
for an anisotropic, harmonically confined QD, is written as
\begin{equation}
H({\bf r}) = T + \frac{1}{2} m^* (\omega^2_{x} x^2 + \omega^2_{y} y^2),
\label{hsp}
\end{equation}
where $T=({\bf p}-e{\bf A}/c)^2/2m^*$, with ${\bf A}=0.5(-By,Bx,0)$ being the
vector potential in the symmetric gauge. The effective mass is $m^*=0.07m_0$, and
${\bf p}$ is the linear momentum of the electron. The second term is the
external confining potential. In the Hamiltonian (\ref{hsp}), we neglect the
Zeeman contribution due to the negligible value ($g^* \approx 0$) of the
effective Land\'{e} factor in our sample.

In the EXD method, the many-body wave function is written as a linear
superposition over the basis of non-interacting 2$e$ determinants, i.e.,
\begin{equation}
\Psi^{s,t}_{\text{EXD}} ({\bf r}_1, {\bf r}_2) =
\sum_{i < j}^{2K} \Omega_{ij}^{s,t} | \psi(1;i) \psi(2;j)\rangle,
\label{wfexd}
\end{equation}
where $\psi(1;i) = \varphi_i({\bf r}_1)\alpha(1)$ if $1 \leq i \leq K$ and
$\psi(1;i) = \varphi_{i-K}({\bf r}_1)\beta(1)$ if $K+1 \leq i \leq 2K$ [and
similarly for $\psi(2,j)$], with $\alpha(\beta)$ denoting up (down) spins;
the index $i$ (or $j$) $\equiv (k,l)$ and $\varphi_i ({\bf r}) = X_k(x) Y_l(y)$,
with $X_k(Y_l)$ being the eigenfunctions of the one-dimensional oscillator in the
$x$($y$) direction with frequency $\omega_x$($\omega_y$). 
In the calculations we use $K=79$, yielding convergent results.
The total energies $E^{s,t}_{\text{EXD}}$ and the coefficients
$\Omega_{ij}^{s,t}$ are obtained through a ``brute force'' diagonalization of
the matrix eigenvalue equation corresponding to the Hamiltonian in Eq.\
(\ref{ham}). 

\begin{figure}[tbhp]
\centering
\includegraphics[width=6.5cm]{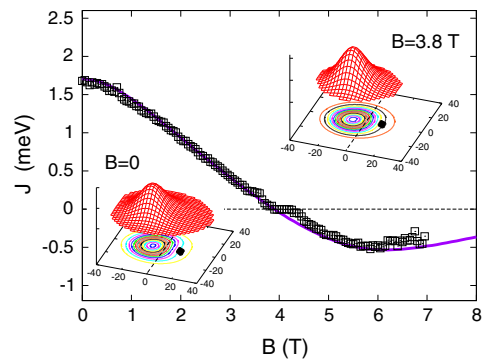}
\caption{(Color online) The ST splitting $J(B)$ extracted from the experiment 
(open squares) together with the calculated EXD results (solid line). For the 
dot parameters, see text. Inset: Conditional probability distributions for the 
second electron when the first is held fixed at the point indicated by the solid
dot. For the $2e$ QD that we studied, the Wigner molecule character is
gradually enhanced with increasing magnetic field. Lengths in nm. 
CPDs in arbitrary units.
\label{figure3}
}
\end{figure}

Figure~\ref{figure3} shows the experimentally extracted energy splitting $J(B)$ 
between the states $S_0$ and $T_+$ together with results of the EXD calculations
with $\hbar \omega_x=4.23$ meV, $\hbar \omega_y=5.84$ meV [i.e., 
$\hbar \omega_0=\hbar \sqrt{(\omega_x^2+\omega_y^2)/2}=5.1$ meV], and
$\gamma=0.862$ \cite{note1}.
With this value of $\gamma$, the Wigner parameter $R_\mathrm{W}=1.34$. 
The anisotropy used in the calculations is $\omega_y/\omega_x=1.38$ indicating that the quantum dot considered here is closer to being circular than in other experimental systems \cite{Kyriakidis02,Zumbuhl04}.

To assess the dot parameters determined from the comparison between the
measured and calculated $J(B)$, we checked that varying the value of $\omega_0$ 
(while keeping $\omega_y/\omega_x=1.38$) and the parameter $\gamma$ does not
influence substantially the quality of the fit to the experimental $J(B)$ $-$
as long as $\hbar \omega_0$ does not deviate greatly (i.e., more than 20\%) from
the $N=1$ single-particle excitation $\Delta_1$. Such range of variations in the
dot parameters is reasonable for the given sample and its measured $N=1$ 
excitation and charging energies and is due to changes in the confinement with 
changing $V_\mathrm{pg}$.

The inset of Fig.~\ref{figure3} shows the EXD calculated singlet conditional 
probability distributions (CPDs) \cite{Yannouleas00,mikh,maks,yann04,note68}
for the second electron, given that the first 
electron is kept fixed at a given location (indicated by a solid dot). The 
electrons clearly tend to avoid each other reflecting their strongly correlated 
motion, in contrast to the two electrons in the natural Helium  atom that
occupy the same single-particle orbital. 
Comparison of the CPDs for $B=0$ and $B=3.8$ T shows that the magnetic field 
leads to an enhancement of the correlation between the two electrons resulting 
in a sharpening of the Wigner-molecular state. Larger values of $B$ result in an
effective dissociation \cite{yann7} of the WM in analogy to the dissociation of
the natural H$_2$ \cite{ref2}, leading eventually to a vanishing 
of $J(B)$, in agreement with the trend seen in Fig.~\ref{figure2}. 
We emphasize that
the negligible Zeeman splitting in our sample is crucial for the observation of this trend  which should even lead to ST oscillations of the ground state spins at higher magnetic fields \cite{Wagner92}. Experimentally, however, the accessible magnetic field range is limited, because conductance resonances disappear with increasing field due to wave function shrinkage, and counterbalancing this effect by changing gate voltages would seriously change the confinement potential.

The EXD calculated spectrum of the elliptic $2e$ QD is displayed in Fig.\ 4.
The colored curves correspond to features highlighted in the experimental
spectrum (see Fig.\ 2) in the region between $N=1$ and
$N=2$. The good overall agreement between the measured and calculated spectra 
allows for a full interpretation of the observed features and their physical
origins rooted in strong electronic correlations. The energy difference
between the $T_+$ and $S_0$ states and its variation with $B$ were discussed 
above [see $J(B)$ in Fig.\ 3].    

Further interpretation is facilitated by consideration of the EXD calculated
spectrum of the corresponding circular dot (Fig.\ 4, inset) where all the
many-body energy levels $(0,m)$ with zero nodes and angular momentum index 
$m > 0$ decrease in energy with increasing magnetic field, converging at much 
higher $B$ to the lowest Landau level of the interacting electrons. Since the
deformation of the elliptic QD discussed here is moderate, we adopt in Fig.\ 4
the same notation as for the circular dot (see caption of Fig.\ 4).
The second excited singlet state seen in the experiment (Fig.\ 2) and its avoided
crossing with the first excited singlet state at $B \approx 5.8$ T is well 
reproduced by the theory [see curves marked as $S_2$ (red) and $S_0$ (magenta) in
Fig.\ 4; the avoided crossing is marked by a double arrow] \cite{note23}.
The third excited state, $T_{+,CM}$, is identified as a (1,0,0,1) triplet 
excited state (green online) that is related to the $T_+$ state (blue online) 
through a CM excitation along the $x$ direction. 
We note here that the aforementioned avoided crossing between the $S_0$ 
and the $S_2$ excited states developed as a result of the shape deformation
of the dot and out of the crossing between the (0,0) and (0,2) singlets of the
circular dot (marked by a circle in the inset). Another correspondence between
the measured and calculated spectra is the splitting at $B=0$ between the
$T_+$ and $T_-$ states (measured as 1.5 meV and calculated to be 1.9 meV);
these triplet states are degenerate for the circular dot (see inset).

\begin{figure}[tbhp]
\centering
\includegraphics[width=6.5cm]{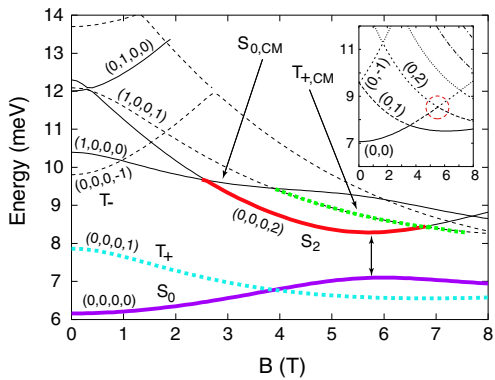}
\caption{(Color online) Calculated EXD energy spectrum, referenced to 
$2 \hbar \sqrt{\omega_0^2 + \omega_c^2/4}$, in a magnetic field of a 
2$e$ dot with anisotropic harmonic confinement 
(for the dot parameters, see text). 
We have adopted the notation $(N_x,N_y,n,m)$, where $(N_x,N_y)$ refer to the
CM motion along the $x$ and $y$ axes and $(n,m)$ refer to the number of radial 
nodes and angular momentum of the relative motion in the corresponding 
{\it circular\/} dot.
Inset: The EXD spectrum of the corresponding circular dot. Only the $(n,m)$ 
indices are shown, since $N_x=N_y=0$ for all the plotted curves. 
Solid lines denote singlets. Dashed lines denote triplets.
\label{figure4}
}
\end{figure}

In conclusion, we have presented measurements of the excitation spectra of a 
lateral $2e$ QD, with negligible Zeeman splitting, moderate anisotropy and parabolic $z$-confinement being the main differences to systems investigated before. Exact-diagonalization calculations,
including a rather small (14\%) decrease in the effective interelectron Coulomb
repulsion, explain in detail the observed features of the excitation spectra
under the influence of a variable magnetic field, including the variation of the
ST splitting, $J$, as a function of $B$. Theoretical analysis of the
salient excitation-spectra patterns (e.g., avoided level crossings, the line 
shape of $J(B)$, and the lifting of level degeneracies in comparison to the 
circular case) indicates significant electronic correlations, indicating the
formation of a well developed Wigner molecule in an anisotropic quantum dot.
In particular, the behavior of $J(B)$ for larger fields suggests an analogy 
between the dissociation process of the WM and that of the natural H$_2$
molecule. These results are expected to assist the design and implementation 
of quantum-dot based devices, as well as to further advance our understanding
of the interplay between the confinement strength, shape anisotropy, and
degree of screening that also characterize other recent measurements on 
few-electron quantum dots \cite{Kouwenhoven97,Kyriakidis02,Zumbuhl04}.

We thank R. Nazmitdinov for valuable discussions.
Financial support by the Swiss Science Foundation (Schweizerischer
Nationalfonds) is gratefully acknowledged. C.Y. and U.L. acknowledge support
from the US D.O.E. (Grant No. FG05-86ER45234) and the NSF 
(Grant No. DMR-0205328).


\begin{thebibliography}{100}
\bibitem{note23}
The larger calculated energy gap at the avoided crossing in comparison to
the measured value may be due to deviations from a harmonic confinement.
\end{thebibliography}
\end{document}